\documentclass[jcp,twocolumn,showpacs,preprintnumbers,superscriptaddress]{revtex4-1}
 
\usepackage{amsfonts}
\usepackage{graphicx}
\usepackage{amsmath}
\usepackage{amssymb}
\usepackage{latexsym}
\usepackage{psfrag}

\usepackage{dcolumn}
 \usepackage{datetime}
\usepackage{multirow}

\usepackage{subfigure}
\usepackage[bookmarks, colorlinks=true, plainpages = true, linkcolor = blue, citecolor = blue, urlcolor = blue, filecolor = blue]{hyperref}

\begin{document}
\title{Identification of the Atomic Scale Structures of the Gold-Thiol Interfaces of Molecular Nanowires by Inelastic Tunneling Spectroscopy}
 
\author{Firuz Demir} 
\email[Email: ]{fda3@sfu.ca}
\affiliation{Department of Physics, Simon Fraser
University, Burnaby, British Columbia, Canada V5A 1S6}

\author{George Kirczenow} 
\email[Email: ]{kirczeno@sfu.ca}
\altaffiliation{Canadian Institute for Advanced Research, Nanoelectronics Program.}
\affiliation{Department of Physics, Simon Fraser
University, Burnaby, British Columbia, Canada V5A 1S6}

\date{\today}

\begin{abstract}\noindent

We examine theoretically the effects of the bonding geometries at the gold-thiol interfaces on the inelastic tunneling spectra of propanedithiolate (PDT) molecules bridging gold electrodes and show that inelastic tunneling spectroscopy combined with theory can be used to determine these bonding geometries experimentally. With the help of density functional theory, we calculate the relaxed geometries and vibrational modes of extended molecules each consisting of one or two PDT molecules connecting two gold nanoclusters. We formulate a perturbative theory of inelastic tunneling through molecules bridging metal contacts in terms of {\em elastic} transmission amplitudes, and use this theory to calculate the inelastic tunneling spectra of the gold-PDT-gold extended molecules. We consider PDT molecules with both trans and gauche conformations bound to the gold clusters at top, bridge and hollow bonding sites. Comparing our results with the experimental data of Hihath {\em et al.}  [Nano Lett. \textbf{8 }, 1673 (2008)], we identify the most frequently realized conformation in the experiment as that of trans molecules top-site bonded to both electrodes. We find the switching from the 42~meV vibrational mode to the 46~meV mode observed in the experiment to be due to the transition of trans molecules from mixed top-bridge to pure top-site bonding geometries. Our results also indicate that gauche molecular conformations and hollow site bonding did not contribute significantly to the experimental inelastic tunneling spectra. For pairs of PDT molecules connecting the gold electrodes in parallel we find total elastic conductances close to twice those of single molecules bridging the contacts with similar bonding conformations and small splittings of the vibrational mode energies for the modes that are the most sensitive to the molecule-electrode bonding geometries. 
\end{abstract}

 
\maketitle

\section{Introduction}
Electrical conduction through single organic molecules bridging a pair of
metal electrodes has been studied extensively, both experimentally and theoretically for more than a decade \cite{review2010}. Knowledge of the microscopic details of the bonding between the molecule and the electrodes is required in order to understand the transport properties of these systems. For molecules amine-linked to gold electrodes, it has been argued\cite{Hybertsen} that only one bonding geometry should be important. However, for the more commonly studied molecular wires with gold-thiolate bonding, it has been suggested that a variety of molecule-metal interface geometries may occur and influence electrical conduction through these
nanowires\cite{review2010,expt_stat_STM_2003,Xiao2004, theo_stat_STM_2005, Li2006, Andrews2008}. Some possible geometries that have been proposed are those in which a sulfur atom of the molecule is located at a {\em top} site over a particular gold surface atom or over a {\em bridge} site between two atoms of the gold surface or over a {\em hollow} site of between three gold atoms.\cite{Sellers93, Gronbeck2000, Bilic05, Vericat10} 
However, which (if any) of these possibilities are actually realized in metal-molecule-metal nanowires has remained unclear\cite{review2010}. Since a single molecule located between two electrodes and chemically bonded to both of them is not accessible to scanning microprobes that are sensitive to atomic-scale structure, direct  experimental determination of the atomic scale geometries of the molecule-metal interfaces of metal-molecule-metal nanowires has been elusive and continues to be an important goal in this field \cite{review2010}.  
 In this paper, we show how inelastic tunneling spectroscopy (IETS) combined with {\em ab initio} computer simulations can distinguish between the different possible atomic-scale structures of the gold-thiolate interfaces of single-molecule nanowires and can therefore be used to definitively identify the molecule-electrode bonding geometries that are realized experimentally. A brief account of some of our preliminary results has already been published \cite{Demir2011}.
 
When the bias voltage applied across a molecular wire increases beyond the threshold value for the excitation of a vibrational quantum of the wire (to be referred to as a phonon) at low temperatures, a conductance step is observed in the  current-voltage characteristic of the system.  
IETS experiments have measured the molecular wire phonon energies in this way for various molecules.\cite{
Park2000, Kushmerick2004, Qiu2004, WangLeeKretzschmarReed04, Yu2004, Cai2005, Djukic2005, 
Wang2006, Kushmerick2006, Long2006, Naydenov2006, Parks2007, Yu2007, LongTroisi07, Wang2008, deLeon2008,
HihathArroyoRubio-BollingerTao08, KiguchiTalWPKDCvanRuitenbeek08, Song2009, Rahimi2009, Song2009a, Taniguchi2009, Tsutsui2009, Okabayashi2010, Song2010, Arroyo2010, Kim2011, Secker2011, KimarXiv, Kimarstretch, Song2011} Theoretical studies of the 
IETS of molecular nanowires \cite{Bonca9597, Emberly2000, May2002, Chen2003, Seideman2003, 
TroisiRatnerNitzan03, Cizek2004, Galperin20040707, Galperin2008, Pecchia2004, 
Pecchia2007, Sergueev2005,TroisiRatner05, Troisi2006, Troisi2006JCP, 
Benesch2006, Jean2006, Paulsson2006, Paulsson2008, Yan2006,
Walczak200607, Frederiksen2007, LaMagna2007, TroisiRatner07, 
Gagliard2007, TroisiBeebe2007, Hartle2008, Kula2008, Luffe2008, 
Jiang2008, Troisi2008, Shimazaki2008, Paulsson.C.F.B.2009, Demir2011, Lin2011bonding,  Lin2011angle, Ueba2007} have also been carried out and
have accounted for many aspects of the IETS data.
These IETS experiments  and comparison with the theoretical work have demonstrated
conclusively that particular molecular species are involved in electrical conduction through metal-molecule-metal junctions by exploiting the fact that the measured phonon energies provide a fingerprint of the molecule involved in the conduction process. The effects of changes in the molecular conformation \cite{Rahimi2009, Paulsson.C.F.B.2009, KimarXiv} and in the orientation of molecules relative to the electrodes\cite{TroisiRatner07, Lin2011angle} on the inelastic tunneling spectra have also been studied. However, the possibility that IETS might identify the {\em bonding} geometries at the molecule-metal {\em interfaces} and thus resolve the long standing ``contact problem" of determining the atomic scale structure of the molecule-electrode interfaces has only very recently begun to be explored.\cite{Demir2011, Lin2011bonding}.

We consider one of the simplest organic molecules, 1,3-propanedithiolate~(PDT), bridging gold
electrodes. Electrical conduction through this molecular nanowire and its IETS have recently been studied experimentally by Hihath \emph{et al.}\cite{HihathArroyoRubio-BollingerTao08} using the statistical STM break junction technique: By repeatedly forming a gold-PDT-gold molecular nanowire, measuring its conductance characteristic and then breaking the wire and reforming it, Hihath \emph{et al.}\cite{HihathArroyoRubio-BollingerTao08} collected a large body of conductance and IETS data representing many separate realizations of this system. However, in common with previous  break junction experiments \cite{review2010},  the specific geometry corresponding to any particular measurement was not identified. 

In the present study, we calculated the relaxed geometries, vibrational normal modes,  zero bias conductances, and IETS of trans and gauche PDT molecules bonded to pairs of gold clusters in top, bridge, and hollow conformations. An important aspect of the methodology that we used is that we carried out relaxations of the {\em entire} extended molecules that we studied, relaxing the positions of {\em all of the atoms of the gold clusters} as well as those of the atoms of the PDT molecule without any imposed constraints. This is necessary in order to obtain reliable results for the vibrational modes that have the largest amplitudes on the sulfur atoms that bond to the gold clusters and are therefore the most sensitive to the sulfur-gold bonding geometry. We also studied the dependence of our results on the sizes of the gold clusters and found good convergence of our results with increasing cluster size. Thus our results are expected to be relevant to PDT molecules bridging macroscopic gold electrodes as in experiments, and we do indeed find excellent agreement between our results and the experimental data of Hihath \emph{et al.}\cite{HihathArroyoRubio-BollingerTao08}. We identify the most frequently realized conformation in the experiment\cite{HihathArroyoRubio-BollingerTao08} as that of trans molecules top-site bonded to both electrodes. We find the switching from the 42~meV vibrational mode to the 46~meV mode observed in the experiment \cite{HihathArroyoRubio-BollingerTao08} to be due to the transition of trans molecules from mixed top-bridge to pure top-site bonding geometries. We also conclude that gauche molecular conformations and hollow site bonding did not contribute significantly to the experimental inelastic tunneling spectra \cite{HihathArroyoRubio-BollingerTao08}. For pairs of PDT molecules connecting the gold clusters in parallel we find total elastic conductances close to twice those of single molecules bridging the contacts with similar bonding conformations and small splittings of the vibrational mode energy of interest.

The remainder of this paper is organized as follows: 
In Section~\ref{DFT_ExtendedMolecule}, we outline how we calculate the equilibrium geometries and vibrational modes for the gold-PDT-gold molecular wires. 
In Section~\ref{elastic_transport}, we describe how the low bias elastic tunneling conductances of these molecular wires are calculated. 
In Section~\ref{perttheory}, we derive the perturbative expression that we use to calculate the conductance step heights associated with inelastic electron scattering in the molecular wires. 
In Section~\ref{relaxation}, we discuss the relaxed molecular wire geometries that we find. 
In Section~\ref{ETSresults}, we present the results of our low bias elastic conductance calculations. 
In Section~\ref{IETSresults}, we present the results of our calculations of the inelastic tunneling spectra of PDT molecules connecting gold electrodes and compare them with the experimental measurements of Hihath \emph{et al.}\cite{HihathArroyoRubio-BollingerTao08} 
We present our conclusions in Section~\ref{Conclusions}.

\section{Theory}
\subsection{Calculation of the low energy conformations and vibrational normal modes}
\label{DFT_ExtendedMolecule}
Our goal is to study inelastic tunneling processes that are sensitive to the structure of the gold-thiol interfaces. These processes typically involve  excitation of vibrational modes with strong amplitudes on the sulfur atoms. Therefore, it is necessary to calculate accurate equilibrium geometries of the entire system being studied, including both the molecule and the gold electrodes, and the phonon energies and atomic displacements from equilibrium for the vibrational normal modes of the {\em whole} system.
We do this by performing our calculations for  {\em extended} molecules that consist of the molecule itself and two clusters of gold atoms (with up to 14 gold atoms per cluster) that represent the pair of electrodes to which the molecule binds. 

In order for this to be practical, we perform {\em ab initio} relaxations of these extended molecules based on density functional theory~(DFT)\cite{HohenbergKohn64, KohnSham65} with the GAUSSIAN~09 \cite{Gaussian} package. We obtain relaxed structures for the various different bonding geometries and molecular conformations that we study by carrying out relaxations starting from different initial trial geometries. The relaxations are carried out with {\em no constraints imposed on the geometries} so that the relaxed geometries correspond to local minima of the total energy and therefore all of the computed vibrational modes about the relaxed geometries are well defined.
 
We note that in most previous calculations of the inelastic tunneling spectra of molecules bridging gold electrodes some or all of the gold atoms were frozen in the geometry of an ideal bulk gold crystal lattice. In that case the calculated geometries of the extended molecules were only partly relaxed and did not correspond to minima of the total energy. Therefore calculations  of vibrational modes based on these partly relaxed geometries suffered from a lack of rigor. Furthermore in the experiment of Hihath \emph{et al.}\cite{HihathArroyoRubio-BollingerTao08} the molecular junctions were formed by crashing a gold STM tip into a gold substrate and then breaking the junction and repeating this process many times at cryogenic temperatures. This experimental ``statistical STM break junction" methodology makes it very unlikely that the gold has the structure of an ideal bulk gold crystal anywhere near the PDT molecule that bridges the gold electrodes in the experiment. Thus models in which some of the gold atoms are frozen in the geometry of an ideal bulk gold crystal lattice are not expected to be appropriate for experiments of this type. This may explain why the authors of Ref. \onlinecite{Lin2011bonding} had difficulty modeling the results of IETS experiments on molecular junctions formed by the mechanically controlled break junction technique.      

In the present work, the atomic displacements from equilibrium in the normal modes and the corresponding frequencies and phonon energies are also calculated using DFT  \cite{Gaussian}. By carrying out systematic calculations for gold clusters of different sizes, we have checked that our conclusions are independent of the cluster size for the larger clusters that we study and thus are applicable to molecules bridging the nanoscale tips of experimentally realized macroscopic gold electrodes.

We use the B3PW91 density functional (for the exchange-correlation energy) and the Lanl2DZ pseudo-potentials and basis sets.\cite{Gaussian,PerdewBurkeErnzerhof1996} However, we have repeated the calculations for some cases using the functional PBE0(PBE1PBE)\cite{Gaussian,PerdewBurkeErnzerhof1997} for comparison and found little change in the result.
We calculated the vibrational frequencies without using any scaling factors and will, in what follows, compare the corresponding calculated phonon energies with the experimental data of Hihath \emph{et al.}\cite{HihathArroyoRubio-BollingerTao08} without making any adjustments. 

\subsection{Calculations of the elastic conductances in the limit of low bias}  
\label{elastic_transport}

Conduction through molecular nanowires in the low bias limit does not involve inelastic processes. 
We calculate the low bias conductances in this limit for the {\em ab initio} extended molecule geometries obtained as described in Section \ref{DFT_ExtendedMolecule}  using  the extended H\"{u}ckel model with the parameters of Ammeter {\em et al.}\cite{yaehmop}  to estimate the electronic structures for these geometries. 
As is discussed in detail in Refs.~\onlinecite{review2010} and \onlinecite{Cardamone08},
this methodology involves no fitting to any experimental data relating to transport in molecular wires or PDT molecules with gold contacts. Previous transport calculations based on extended H\"{u}ckel  theory have yielded elastic tunneling conductances in agreement with experiment for other molecules thiol bonded to gold electrodes \cite{Datta1997, EmberlyKirczenow01, EmberlyKirczenow01PRB, Kushmerick02, Cardamone08} 
and have also explained a number of transport phenomena observed in experiments on molecular arrays on silicon\cite{PivaWolkowKirczenow05, PivaWolkowKirczenow08, PivaWolkowKirczenow09} and graphene nanoribbons\cite{Ihnatsenka} as well as electroluminescence data,\cite{Buker08} current-voltage characteristics\cite{Buker08} and STM images\cite{Buker05} of molecules on complex substrates. Our calculated elastic tunneling conductances (at zero bias) for the gold-PDT-gold molecular nanowires are in good agreement with the values measured by Hihath \emph{et al.}\cite{HihathArroyoRubio-BollingerTao08} 

We calculate the zero bias tunneling conductances $g$ for gold-PDT-gold molecular
wires from the Landauer formula 
\begin{equation}\label{tansm } 
{\small 
g = g_0 \sum_{ij}    
\Big\vert t_{ji}^{el}(\{{\bf 0}\}) \Big\vert ^2 
\frac{v_j}{v_i} 
}~,
\end{equation}
where $g_0=2e^2/h$ and $t_{ji}^{el}(\{{\bf 0}\})$ is the
elastic transmission amplitude through the extended molecule at the Fermi energy 
$E_F$ of bulk
gold calculated within extended H\"{u}ckel theory using the standard parameter set of Ammeter {\em et al.}\cite{yaehmop}
In the transmission
amplitude, $\{{\bf 0}\}$ means that the extended molecule is in a relaxed geometry
calculated as is discussed in Sec. \ref{DFT_ExtendedMolecule}, 
$i$ denotes an electronic state of a carrier that is coming from the left lead, and
$j$ is the electronic state of a carrier that has been transmitted to the right lead. 
$v_j$ and $v_i$ are the corresponding electron velocities.
The coupling of the extended molecule to the electron reservoirs was treated as in previous work
\cite{PivaWolkowKirczenow05, DalgleishKirczenow05, DalgleishKirczenow06, Kirczenow07, PivaWolkowKirczenow08, PivaWolkowKirczenow09, Cardamone08, Cardamone10, SMM}
by attaching a large number of semi-infinite quasi-one-dimensional ideal leads to the valence orbitals of the outer gold atoms of the extended molecule. This method of coupling the extended molecule to the reservoirs is successful because most of the resistance of this system is due to the weakly transmitting molecule itself (as distinct from the coupling between the extended molecule and the ideal leads) and because the many ideal leads behave similarly to a large number of phase-randomizing B\"{u}ttiker probes\cite{Buttikerprobes} in minimizing the influence of electronic dimensional resonances due to the finite sizes of the gold clusters of the extended molecule. 

The transmission amplitudes $t_{ji}$ were found by solving the Lippmann-Schwinger equation 
\begin{equation}\label{Lippmann-Schwinger} 
\vert \Psi \rangle =  \vert \Phi_0 \rangle +
G_0(E) W  \vert \Psi \rangle ~,
\end{equation}
where $ \vert \Phi_0 \rangle$ is an electron eigenstate of 
an ideal semi-infinite one dimensional left lead that is decoupled from the extended molecule,  
$G_0(E)$ is the Green's function of the decoupled system of the ideal leads 
and the extended molecule, 
$W$ is the coupling between the extended molecule and leads, 
and $\vert \Psi \rangle$ is the scattering eigenstate of the complete coupled system. 
Since the basis set used in extended H\"{u}ckel theory is non-orthogonal, the orthogonalization procedure introduced in Ref.~\onlinecite{EmberlyKirczenow98} was used to facilitate the above calculations. 

\subsection{Perturbative theory of the IETS of molecular wires}  \label{perttheory}
When the bias voltage $V$ applied across a molecular wire  exceeds  $\hbar \omega/e$ (where $\omega$ is the frequency of a vibrational mode of the system), it becomes energetically possible for electrons passing through the molecular wire to emit phonons of that mode and a conductance step develops in the low temperature current voltage characteristic of the wire. Since the bias voltages involved are small,  the mode frequencies that we calculate at zero bias using the {\em ab initio} approach described in Section \ref{DFT_ExtendedMolecule} are appropriate for estimating (via $|V|= \hbar \omega/e$) the values of the bias voltage at which conductance steps due to inelastic processes occur. 

In order to estimate the {\em heights} of these conductance steps, 
we adopt a perturbative approach in the spirit of that proposed by 
Troisi \emph{et al.}\cite{TroisiRatnerNitzan03} who transformed the problem of calculating IETS
intensities into an {\em elastic} scattering problem. However, unlike Troisi \emph{et al.}\cite{TroisiRatnerNitzan03}, we formulate the present theory and results explicitly in terms of elastic electron transmission amplitudes $ t_{ji}^{el}$ through the molecular wire. Such perturbative theories are appropriate at low bias voltages provided that the elastic electron tunneling channel is well off 
resonance,\cite{TroisiRatnerNitzan03} conditions that are satisfied for gold-PDT-gold molecular wires in the experiments of Hihath \emph{et al.}\cite{HihathArroyoRubio-BollingerTao08} 
Since we are concerned with the case of weak (far off resonance) electron transmission through the molecule, inelastic forward scattering of electrons through the molecule is expected to make the dominant contribution to the steps in the conductance that are associated with the emission of phonons and the conductance is expected to increase with increasing bias at these steps.\cite{Vega2006, Ueba2007, Paulsson2008} We shall assume this to be the case in the following
derivation of the theory of IETS on which our numerical calculations of IETS intensities presented in Section \ref{IETSresults} are based.

In the harmonic approximation the vibrational Hamiltonian of an extended molecule is 
\[
H_{vib} =\sum_{in} \frac{p_{in}^{2}}{2 m_{n}}  +  
\frac{1}{2} \sum_{in jm}   V_{in jm}  q_{in}  q_{jm}  ~,
\]
where $p_{in}$ and $q_{in}$ are the $i^{th}$ atomic momentum and position coordinate operators for atom $n$. They obey the commutation rules
$[q_{in} , p_{jm}] =i \hbar \delta_{i,j}\delta_{n,m}$. The position coordinate $q_{in}$ is measured
from the equilibrium position of the atom $n$.

In terms of the creation and destruction operators $a_{\alpha}^{\dagger}$ and $a_{\alpha}$
for phonons of the normal modes $\alpha$ of the extended molecule with frequencies $\omega_\alpha$, the harmonic Hamiltonian takes the form
\[
H_{vib} =\sum_{\alpha} \hbar \omega_\alpha
\left(
a_{\alpha}^{\dagger} ~a_{\alpha}^{} + \frac{1}{2} 
\right)   ~
\]
and the atomic position coordinate operators are 
 {\small
\begin{equation}\label{displtophonon} 
q_{in}  = \sum_{\alpha}  \sqrt{ \frac{\hbar}{2 \omega_{\alpha}} }
       (d^*_{{in}\alpha} a_{\alpha}^{} +  d_{{in}\alpha} a_{\alpha}^{\dagger}) ~.
\end{equation}
}Here the coefficients ${\bf d}_{{n}\alpha}$ are the displacements from their equilibrium 
positions of the atoms $n$ of the extended molecule in normal mode 
$\alpha$ normalized so that $\sum_{n}  m_n  {\bf d}^*_{{n}\alpha'}\cdot
{\bf d}_{{n}\alpha} = \delta_{\alpha',\alpha}$ and $m_n$ is the mass of 
atom $n$. 

Let us now approximate the interaction Hamiltonian $U({\bf r})$
describing the interaction between an electron at position ${\bf r}$ and
the extended molecule by a sum of terms
\begin{equation}\label{couplingH} 
U({\bf r}) = \sum_{n}  u_n ({\bf r}-{\bf R}_n) ~, 
\end{equation}
 corresponding to the interactions of the
electron with the individual atoms $n$ of the extended molecule whose nuclei 
are located at ${\bf R}_n$. 
 
Then the change in the Hamiltonian $U({\bf r})$ if the atoms are displaced from their 
equilibrium positions ${\bf R}^{0}_n$ is 

\begin{equation}\label{couplingeph} 
\Delta U({\bf r}) = \sum_{n}  u_n ({\bf r}-{\bf R}_n) -  u_n ({\bf r}-{\bf R}^{0}_n). 
\end{equation}
To leading order in the displacements $q_{in}$ of the atoms from their equilibrium positions, this becomes
\begin{equation}\label{couplingephlin} 
\Delta U(r)  \cong  \sum_{in} - q_{in} \frac{\partial u_n ({\bf r}-{\bf R}^{0}_n)}{\partial{r_i} }.
\end{equation}
Using Eq. (\ref{displtophonon} ), this yields the electron-phonon interaction Hamiltonian $H_{ep} $
{\small \[
 H_{ep}  = \Delta U(r)  \cong   -\sum_{\alpha}  \sqrt{ \frac{\hbar}{2 \omega_{\alpha}} }a_{\alpha}^{\dagger} 
\sum_{in} d_{{in}\alpha}  \frac{\partial u_n ({\bf r}-{\bf R}^{0}_n)}{\partial{r_i} }+\mathrm{h.c.} 
\]   } If $A$ is a suitable scale factor and ${\bf d}_{{n}\alpha}$ is the vector with components $d_{{in}\alpha} $ 
this becomes
{\small \[
 H_{ep}   \cong   \sum_{\alpha}  a_{\alpha}^{\dagger} H_{\alpha} (r)   +\mathrm{h.c.}   ~~~ \mbox{where}
\]  }
{\small  \[
 H_{\alpha} (r)= \sqrt{ \frac{\hbar}{2 \omega_{\alpha}} }
                          \lim_{A\to0} \sum_{n}  \frac{u_n ({\bf r}-({\bf R}^{0}_n + A{\bf d}_{{n}\alpha})) -  u_n ({\bf r}-{\bf R}^{0}_n)}{A} .
 \]  }

For the system of interest in this paper, the electron-phonon interaction $H_{ep}$ is a small perturbation. The operators
$ a_{\alpha}^{\dagger}$ and  $a_{\alpha}$ induce transitions
between states with differing numbers of phonon quanta while $H_{\alpha} (r)$ simultaneously  gives rise to
transmission of electrons through the molecular wire in the inelastic channel with scattering amplitude $t_{ji}^{\alpha}$.
Assuming that the system is cryogenic so that the presence of any ambient phonons may be ignored, 
and that $H_{\alpha}$ is so weak that $t_{ji}^{\alpha}$ is effectively  linear 
in the strength of $H_{\alpha}$, and  that $A$ is 
small enough that  
 {\small  \[
 H_{\alpha} (r)= \sqrt{ \frac{\hbar}{2 \omega_{\alpha}} }
                           \sum_{n}  \frac{u_n ({\bf r}-({\bf R}^{0}_n + A{\bf d}_{{n}\alpha})) -  u_n ({\bf r}-{\bf R}^{0}_n)}{A}
 \]  }
 to a good approximation, it follows that
 
 \begin{equation}\label{scalefactors} 
 t_{ji}^{\alpha} \cong  \sqrt{ \frac{\hbar}{2 \omega_{\alpha}}}  \frac{1}{A}\tilde{t}_{ji}^{\alpha},
 \end{equation}
 where $\tilde{t}_{ji}^{\alpha}$ is the inelastic electron transmission
 amplitude that would be found if $H_{\alpha}$ were
 \[
 \tilde{H}_{\alpha} (r)= 
                           \sum_{n} u_n ({\bf r}-({\bf R}^{0}_n + A{\bf d}_{{n}\alpha})) -  u_n ({\bf r}-{\bf R}^{0}_n)~.
 \]

That is, $\tilde{t}_{ji}^{\alpha}$ is the change in the {\em elastic} electron transmission
 amplitude when the perturbation 
$ \tilde{H}_{\alpha}$ is added to the electronic Hamiltonian $H_{equil. ~static}$ of the {\em static molecular wire in its equilibrium geometry}.
But  $H_{equil. static}+ \tilde{H}_{\alpha}$ is simply the electronic Hamiltonian of a static molecular wire whose nuclear positions
${\bf R}_n$ are ${\bf R}_n = {\bf R}^{0}_n + A{\bf d}_{{n}\alpha}$, i.e., they are shifted from their equilibrium positions by vectors 
$A{\bf d}_{{n}\alpha}$ that are proportional to the displacements of the atoms from their equilibrium positions in normal mode $\alpha$.

This means that  
 \[\tilde{t}_{ji}^{\alpha} = t_{ji}^{el}(\{A{\bf d}_{{n}\alpha}\})-t_{ji}^{el}(\{{\bf 0}\}),
 \]
where 
$t_{ji}^{el}(\{A{\bf d}_{{n}\alpha}\})$ is the elastic electron 
transmission amplitude through the molecular wire with each atom $n$
displaced from its equilibrium position by $A{\bf d}_{{n}\alpha}$ where $A$ is
a small parameter and  $t_{ji}^{el}(\{{\bf 0}\})$ is the elastic 
electron transmission amplitude through the molecular wire in its
equilibrium geometry from state $i$ with velocity $v_i$ in the electron source
to state $j$ with velocity $v_j$ in the electron drain. It then follows
from Eq.~\ref {scalefactors} that

 \[ 
 t_{ji}^{\alpha} \cong  \sqrt{ \frac{\hbar}{2 \omega_{\alpha}}}  \lim_{A\to0}  \frac{t_{ji}^{el}(\{A{\bf d}_{{n}\alpha}\})-t_{ji}^{el}(\{{\bf 0}\})}{A}~.
 \]
The analog of Eq. (\ref{tansm }) for the {\em inelastic} electron transmission
 probability associated with the emission of a phonon of mode $\alpha$ is then 
{\small   \[   T_{\alpha}
 = {\displaystyle\sum_{ij}^{} \frac{v_j}{v_i}   \Big\vert t_{ji}^{\alpha}} \Big\vert ^2 
 = \frac{\hbar}{2 \omega_{\alpha}}  \lim_{A\to0} \displaystyle\sum_{ij}^{}  \frac{v_j}{v_i}   
    ~\Big\vert \frac{ t_{ji}^{el}(\{A{\bf d}_{{n}\alpha}\})-t_{ji}^{el}(\{{\bf 0}\})} {A} 
    \Big\vert ^2~.
\]  }
From this it follows immediately that the IETS intensity (i.e., the conductance step height) $\delta g_{\alpha}$ associated with the emission of phonons of mode ${\alpha}$ is
{\small
\begin{equation} \label{omegaIntensityA} 
\begin{split}
\delta g_{\alpha}=& g_{0}~T_{\alpha}~,  \mbox{i.e.,}\\
\delta g_{\alpha}=& \frac{2 e^2}{h} 
   \frac{\hbar}{2 \omega_{\alpha}}  \lim_{A\to0} \displaystyle\sum_{ij}^{}  \frac{v_j}{v_i}   
    ~\Big\vert \frac{ t_{ji}^{el}(\{A{\bf d}_{{n}\alpha}\})-t_{ji}^{el}(\{{\bf 0}\})} { A}
     \Big\vert ^2~,   \mbox{or} \\
\delta g_{\alpha}=&  \frac{e^2}{2\pi \omega_{\alpha}} \lim_{A\to0}\sum_{ij} \frac{v_j}{v_i }   
\Big\vert \frac{ t_{ji}^{el}(\{A{\bf d}_{{n}\alpha}\})-t_{ji}^{el}(\{{\bf 0}\})} { A }
 \Big\vert ^2 ~.
\end{split}
\end{equation}
}

Eq.~(\ref{omegaIntensityA}) expresses the IETS intensities $g_{\alpha}$ explicitly in terms of elastic electron transmission amplitudes  $ t_{ji}^{el}$ through the molecular wire.
It states that, in the leading order of perturbation theory, the scattering amplitude for {\em inelastic} transmission of an electron through the molecular wire is proportional to the change in the {\em elastic} amplitude for transmission through the wire if its atoms are displaced from their equilibrium
positions as they are when vibrational mode $\alpha$ is excited.
We evaluate $ t_{ji}^{el}$ in Eq.~(\ref{omegaIntensityA}) numerically to find the heights $\delta g_{\alpha}$~of the conductance steps that arise from inelastic tunneling processes due to emission of phonons of vibrational mode $\alpha$ by applying the methodology described in Section  \ref{elastic_transport}.
These calculations are carried out at the zero bias Fermi energy since the values of the bias at which the inelastic transmission occurs in the experiments of Hihath \emph{et al.} 
\cite{HihathArroyoRubio-BollingerTao08} are low.

\section{Results}
\subsection{Low energy conformations of the extended molecules}
\label{relaxation}

As has been discussed in Section \ref{DFT_ExtendedMolecule}, we considered extended molecules, i.e., molecules together with finite clusters of gold atoms to which the molecules bond, and relaxed these entire structures (the molecules and gold clusters) using density functional theory-based calculations.\cite{HohenbergKohn64, KohnSham65, PerdewBurkeErnzerhof1996, PerdewBurkeErnzerhof1997, Gaussian} No constraints were applied to the extended molecule geometries during the relaxation. This yielded calculated geometries that correspond to local energy minima of the extended molecules. 

\begin{figure}[t]
\centering
\includegraphics[width=1.0\linewidth]{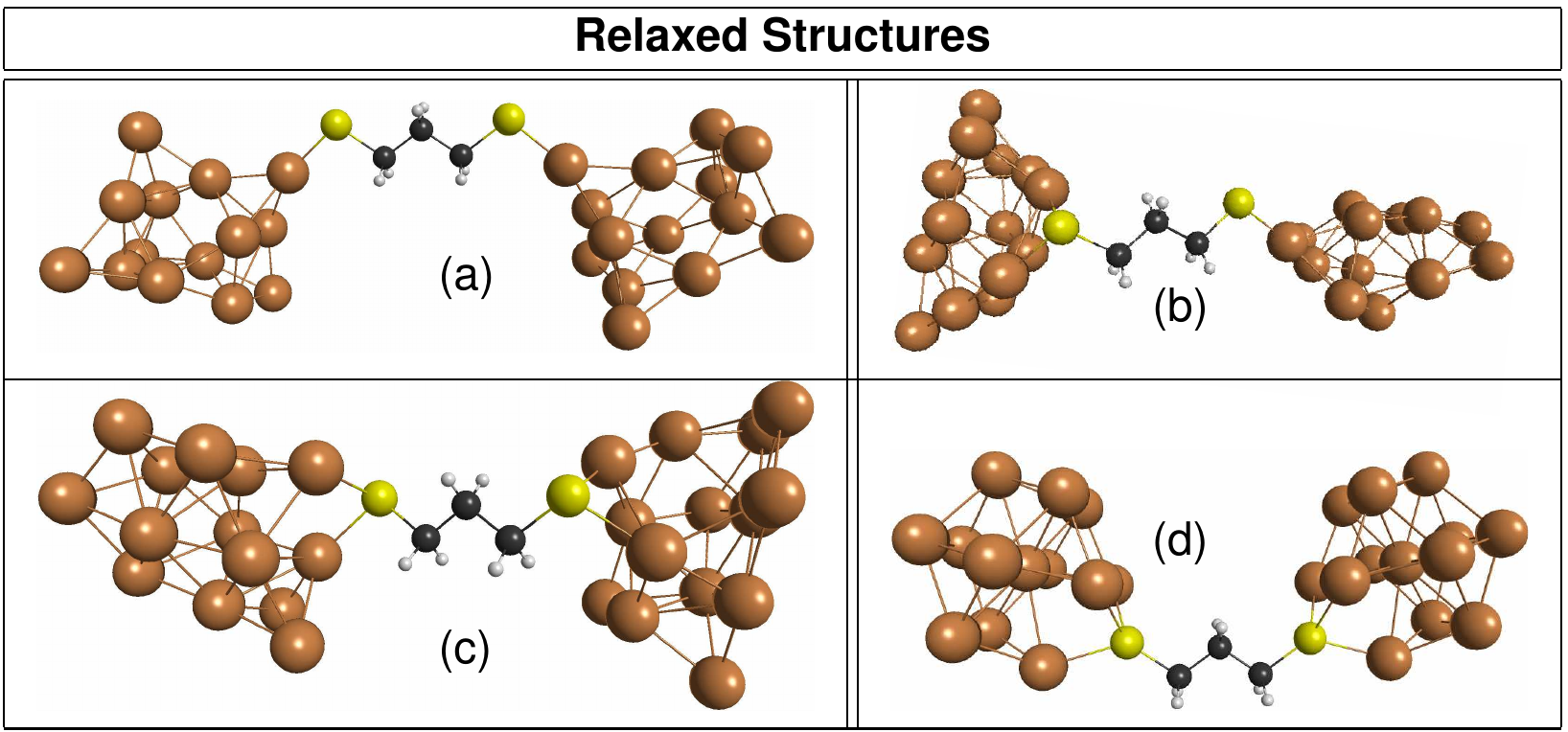}
\caption{{\color{red}{(Color online.) }} Examples of relaxed structures of the trans-PDT molecule bonded to gold clusters in different ways.\cite{Macmolplt}
(a): Both S atoms bond to the gold clusters in the top geometry.  
(b): One S atom bonds to a gold cluster in the top geometry while the other S bonds in the bridge geometry.
(c): Both S atoms bond to the gold clusters  in the bridge geometry.
(d): Both S atoms bond to the gold clusters in the hollow geometry.
Carbon, hydrogen, sulfur, and gold atoms are black, grey,
yellow, and amber, respectively.}
\label{Fig1}
\end{figure}

Representative examples of the relaxed geometries that we obtained for extended molecules in which the PDT molecule has the {\em trans} conformation are shown in Fig.~\ref{Fig1}. 
 Top, bridge and hollow-site geometries where each sulfur atom of the melecule bonds to one, two, or three gold atoms, respectively, are shown in Fig.~\ref{Fig1} (a), (c) and (d). A mixed bridge-top structure where one sulfur atom bonds to two gold atoms and the other bonds to one gold atom is shown in Fig.~\ref{Fig1} (b).

For relaxed extended molecules having structures of the types depicted in Fig.~\ref{Fig1}, we found the extended molecules with the sulfur atoms binding to the
gold clusters in bridge geometries to have the lowest energies. The hollow site bonding geometries had the highest energies. The top site bonding
geometries had energies between those of the bridge and hollow cases. 
For example, the energy of the relaxed extended molecule in the bridge geometry was 
0.46 eV  
lower than that with the top geometry and 
1.33 eV 
lower than that with the hollow geometry for our molecular wires with 13 Au atoms in each gold cluster. 
These numbers varied depending on the gold cluster size. But in all studied cases,  
the bridge geometry was more stable than the top geometry which was more stable than the hollow geometry for relaxed structures. 

We found that  extended molecules for which DFT geometry
relaxations were started with the sulfur atoms over {\em hollow sites} on
the surfaces of close packed gold clusters  having atomic geometries resembling that of bulk fcc gold {\em invariably relaxed to bridge bonding site geometries}.  
This is reasonable physically since the bridge site is not very far from the hollow site and a drastic rearrangement of the gold clusters is not required for a transition from hollow site bonding to a {\em lower energy} bridge site bonding geometry to occur.  
Thus while we were able to generate examples of relaxed extended molecule geometries (such as that in Fig.~\ref{Fig1}~(d)) with each sulfur atom bonding to three gold atoms (i.e., hollow site bonding) the structures of the gold clusters near these bonding sites were much more open than that of a defect-free surface of a fcc gold crystal near a hollow site. Because of the much greater fragility and the higher energies of hollow site bonded structures relative to bridge and top site bonding, it is reasonable to expect hollow site bonding to be realized much less frequently than bridge and top site bonding (if it is realized at all) in statistical STM break junction experiments such as those of Hihath \emph{et al.}\cite{HihathArroyoRubio-BollingerTao08}. As will be seen below, comparison of our theoretical inelastic tunneling spectra with the experimental data of Hihath \emph{et al.}\cite{HihathArroyoRubio-BollingerTao08} supports this expectation.

\subsection{Elastic conductances in the limit of low bias}  \label{ETSresults}
Our calculated zero bias conductance values for
top, bridge, and hollow site bonded {\em trans} PDT molecules are in the ranges 
$g=$0.0012-0.0013$g_0$, 0.0015-0.0019$g_0$, and  0.0032-0.0095$g_0$, respectively, for extended molecules attached to the larger gold clusters that we studied. 
Values for some specific gold cluster sizes are shown in Table \ref{CalculatedConductance}.
{\em All} of these theoretical values agree with the experimental
value $0.006 \pm 0.002 g_0$\cite{HihathArroyoRubio-BollingerTao08} to a degree that is
typical of the experimental and theoretical literature\cite{Lindsay07} for molecules thiol
bonded to gold electrodes. We note that the calculations reported here did not involve 
any fitting to the data of Hihath \emph{et al.}\cite{HihathArroyoRubio-BollingerTao08} or to any other molecular transport experiments.

The calculated conductances show an increasing trend from the top to the bridge to the hollow bonding geometries due to the better electrical coupling between the molecule and electrodes for molecules bonding directly to larger numbers of electrode atoms. The calculated conductances for 
hollow site bonded molecules are closer to the experimental values \cite{HihathArroyoRubio-BollingerTao08} than are the calculated conductances for the the top and bridge site bonding. 
However, as we have already discussed in Section \ref{relaxation}, structures with hollow site bonding are much more fragile and thus are less likely to be realized experimentally than bridge or top site bonded structures. Moreover, as will be shown below in Section \ref{IETSresults}, comparison between our theoretical predictions and the experimental {\em inelastic} tunneling spectra\cite{HihathArroyoRubio-BollingerTao08} provides strong evidence that top and bridge bonding were predominant in the experimentally realized systems of Hihath et al.\cite{HihathArroyoRubio-BollingerTao08}.  Thus it is evident that the present elastic conductance calculations (in common with other elastic conductance calculations for molecules thiol bonded to gold electrodes that are available in the literature \cite{Lindsay07}) are not accurate enough for comparison between the theoretical and experimental {\em elastic} conductance values to reveal which bonding geometries were realized experimentally.     
 
\begin{table}[h!]
\centering
\caption{{\color{red}{(Color online.) }} Calculated IETS intensities and phonon energies for Mode-I for one or two PDT molecules connecting gold clusters together with the calculated low bias conductances. T: top site bonding, B: bridge site bonding, and H: hollow site bonding of a molecule to a gold clusters.  TT and BB indicate two molecules connecting the gold clusters in parallel and bonding in top and bridge geometries, respectively. Gauche1: One gauche C-C bond. Gauche2: Two gauche C-C bonds. Numbers in the configuration column indicate the numbers of atoms in the gold clusters. Notice that the low bias conductance decreases with increasing numbers of gauche bonds for similar bonding geometries, consistent with previous theories for longer molecular wires.\cite{Paulsson.C.F.B.2009,LiPWTBAE2008}
}
\label{CalculatedConductance}
\includegraphics[width=1.0\linewidth]{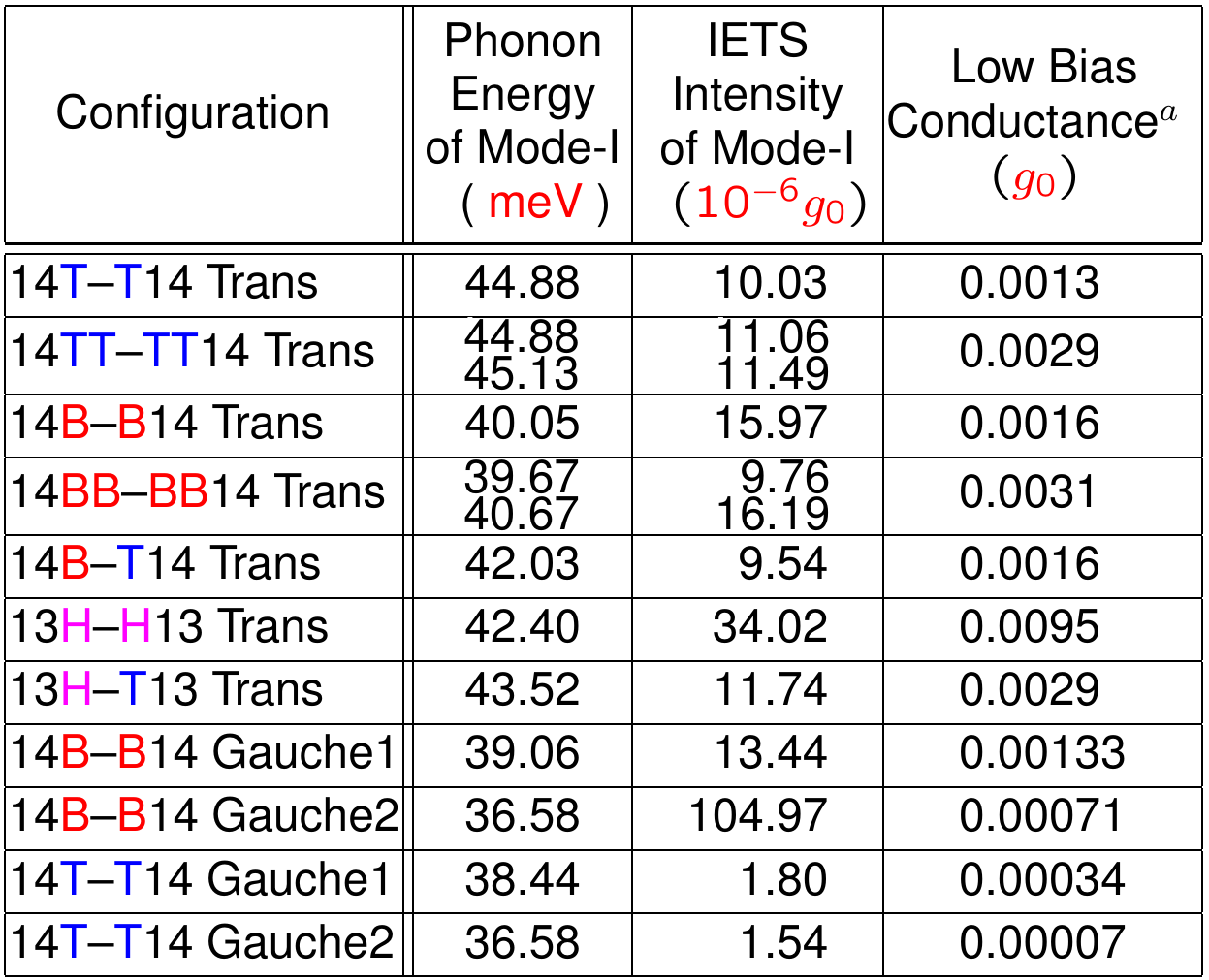}
\footnotetext[1]{Experimental conductance measurement of Hihath \emph{et al.} is about 0.006 $g_{0}$ $\pm 0.002$     \cite{HihathArroyoRubio-BollingerTao08}.    }
\end{table}

\subsection{Inelastic Tunneling Spectroscopy of the Gold-Sulfur Interface}
\label{IETSresults}
In this Section we present the results of our calculations of the inelastic tunneling spectra for trans PDT molecules bonded to gold clusters in top, bridge, and hollow site conformations as well as for molecules with gauche bonds. We also consider some examples of pairs of molecules bridging the gold clusters. We compare our results with the experimental data of Hihath \emph{et al.}\cite{HihathArroyoRubio-BollingerTao08} and deduce which molecular and bonding conformations were most commonly realized in the experiment. In the calculations reported below, density functional theory was used to find the relaxed extended molecule geometries, vibrational modes and their frequencies as described in Sections \ref{DFT_ExtendedMolecule} and 
\ref{relaxation}. The inelastic tunneling intensities $\delta g_{\alpha}$ corresponding to the various modes were calculated from Eq.~(\ref{omegaIntensityA}), evaluating the elastic scattering amplitudes $ t_{ji}^{el}$ that enter Eq.~(\ref{omegaIntensityA}) as is described in Section \ref{elastic_transport}. We will focus primarily on those vibrational modes that have the strongest amplitudes of vibration on sulfur atoms and the strongest IETS intensities since these modes are the most sensitive to the molecule electrode bonding geometries and are also prominent in the experimental IETS spectra. 
\unitlength=0.01\linewidth
\begin{figure}[] 
\centering
\subfigure[] {\label{Fig2(a)} 
\includegraphics[width=1.0\linewidth]{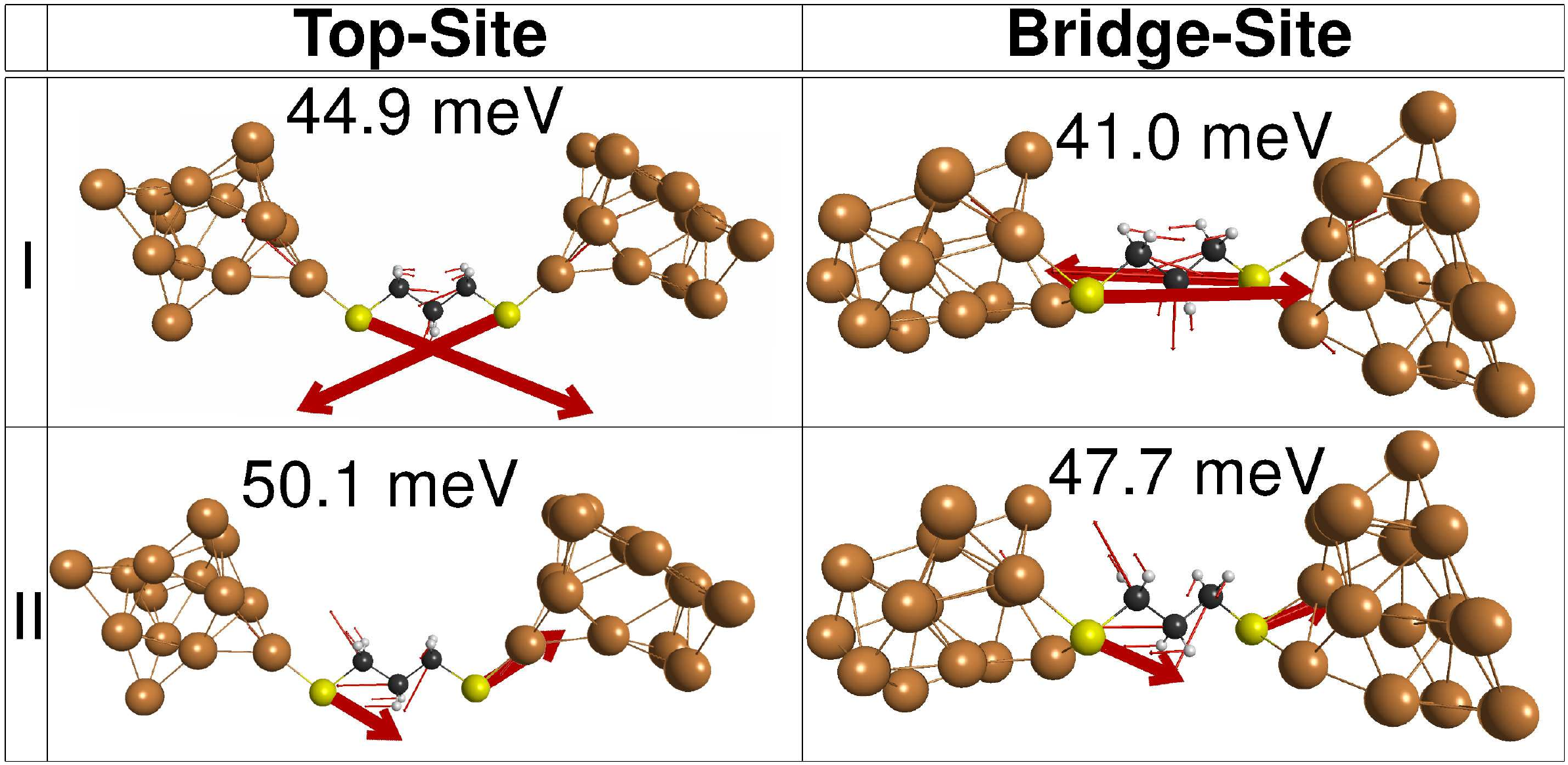}     }
\subfigure[] {\label{Fig2(b)}
\includegraphics[width=1.0\linewidth]{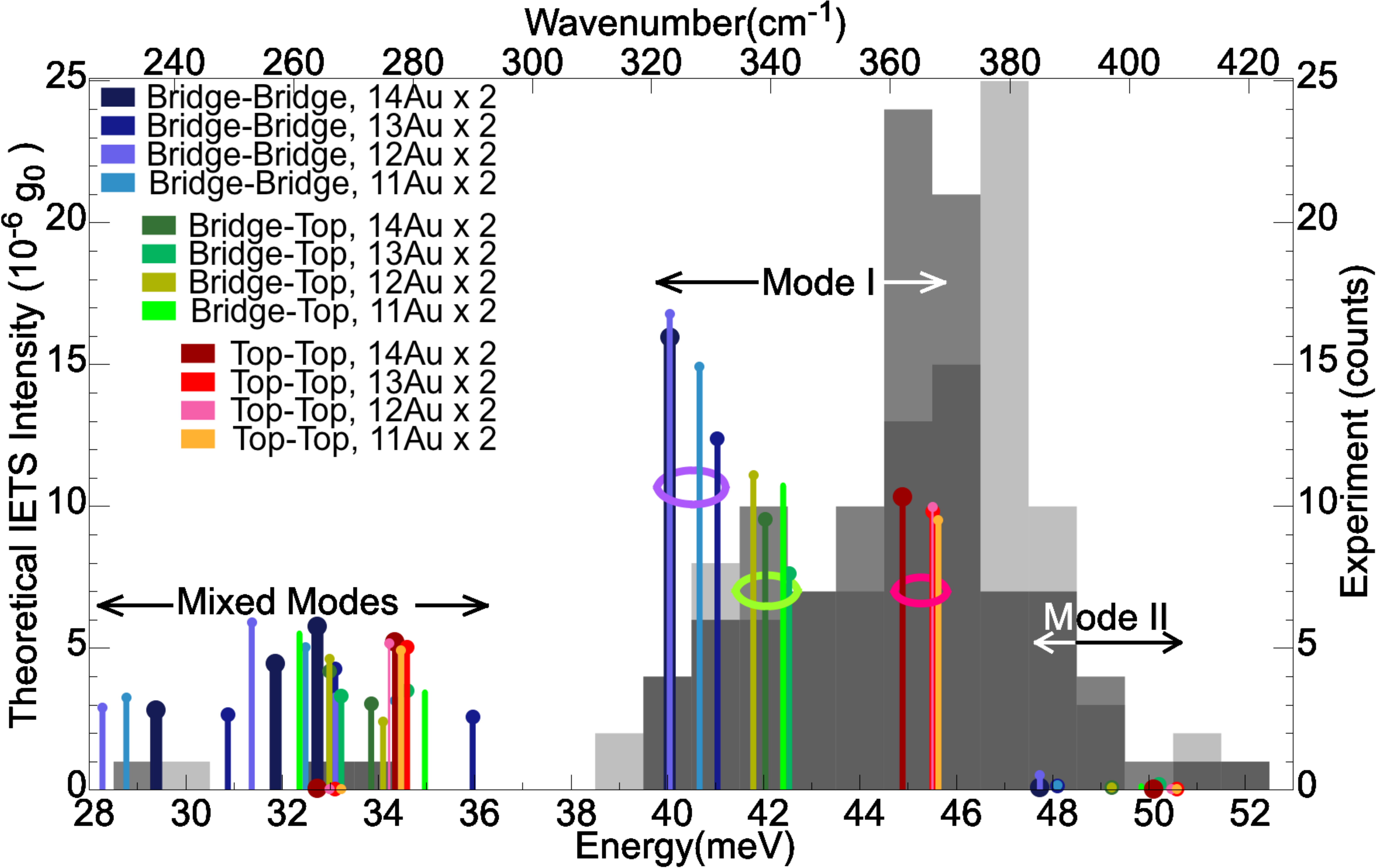}     }
\caption{{\color{red}{(Color online.) }}
\subref{Fig2(a)} 
Calculated vibrational modes in the phonon energy range from $38$ to $53$~meV for {\em trans}-PDT bridging gold nano-clusters with sulfur atoms bonded to gold in top-site and bridge-site geometries. 
Carbon, hydrogen, sulfur, and gold atoms are black, grey, yellow, and amber, respectively.\cite{Macmolplt} 
Red arrows show un-normalized atomic displacements; the heavier arrows indicate the motion of the sulfur atoms. 
Mode-I has the stronger IETS intensities.
\subref{Fig2(b)} 
Calculated IETS intensities (colored) vs. calculated phonon energies for {\em trans}-PDT molecules linking pairs of gold clusters with between 11 and 14 Au atoms in each cluster. 
Results are shown for both sulfur atoms bonding to the gold in top-site and bridge-site geometries and for top site bonding to one gold cluster and bridge site bonding to the other (bridge/top). 
The three ellipses enclose the calculated type I Mode-IETS spectra for pure bridge, pure top, and bridge/top bonding geometries for extended molecules with gold clusters containing various numbers of gold atoms. 
The experimental IETS phonon mode histogram of Hihath \emph{et al.}\cite {HihathArroyoRubio-BollingerTao08} is shown in (darker, lighter) grey for (positive, negative) bias voltages.
Modes of the types I and II in (a) are indicated by arrows. 
}
\end{figure}

\subsubsection{Top and bridge site bonding of trans PDT molecules}
\label{TBHSresults}

Our calculations show that for trans PDT molecules the modes with strong amplitudes of vibration on the sulfur atoms that have the largest IETS intensities fall within the phonon energy range of a prominent feature of the experimental IETS phonon histogram\cite{HihathArroyoRubio-BollingerTao08} (shown in grey in Fig.~\ref{Fig2(b)}) that extends from $39$ to $52$~meV.

The vibrational normal modes in this energy range and their phonon energies are shown in Fig.~\ref{Fig2(a)} for examples of extended PDT molecules with sulfur atoms bonded to 14-atom gold clusters in top and bridge geometries. 
The vibrational modes are labelled I and II according to the nature of the atomic motion. 
The corresponding calculated IETS spectra (IETS intensities vs. phonon energy)  
are shown in color in Fig.~\ref{Fig2(b)} for extended molecules with gold clusters of various sizes  together with the experimental IETS phonon mode histogram\cite {HihathArroyoRubio-BollingerTao08}.

It is important to note that that vibrational modes I and II are strongly localized to the molecule itself and the gold atoms in its immediate vicinity. Their amplitudes of vibration decay rapidly in the gold clusters as the distance from the molecule increases. This is due to the fact that the frequencies of these modes are well above the phonon frequencies of bulk gold so that in the gold these modes are evanescent in character. For the largest gold clusters that we studied, the vibrational amplitudes of modes I and II are very small already before the farthest gold atoms from the molecule are reached. Therefore, the calculated properties of these modes are not impacted significantly by the finite sizes of the gold clusters in the present study.   
 
The mode with the strongest  calculated  IETS intensities in Fig.~\ref{Fig2(b)} is Mode-I that is depicted in the top row of Fig.~\ref{Fig2(a)}. 
In this mode (also known as the Au-S symmetric stretch mode) the sulfur atoms have
the strongest vibrational amplitudes and move in {\em antiphase}, approximately
along the axis of the molecule. 
Mode-II (also known as the Au-S antisymmetric stretch mode) is shown in the lower row of Fig.~\ref{Fig2(a)}. 
It is similar to Mode-I except that in Mode-II the sulfur atoms move {\em in phase} with each other. 
As seen in Fig.~\ref{Fig2(b)} the calculated IETS intensities for Mode-II are much weaker
than those for Mode-I for both top and bridge site bonding.

This difference between the IETS intensities of modes I and II can be understood physically as follows.  
Since in Mode-I the two sulfur atoms move in {\em antiphase} the gold-sulfur distances for both sulfur atoms either increase or decrease {\em together} as the extended molecule vibrates. 
These distances can be regarded as the widths of tunnel barriers between the molecule and the two gold electrodes. 
The motions of the two sulfur atoms act {\em in concert} to widen or narrow {\em both} tunnel barriers together and thus to weaken or strengthen the electron transmission amplitude through the molecular wire. 
Therefore, the magnitude of the difference between the elastic electron transmission amplitudes of equilibrium and vibrating geometries through the molecular wire in Eq.~(\ref{omegaIntensityA}) is enhanced. 
By contrast,  in Mode-II, while the gold-sulfur distance for one sulfur atom increases, the gold-sulfur distance for the  other sulfur atom decreases. 
These effects of the motions of the two sulfur atoms on the elastic transmission amplitude through the molecular wire tend to cancel. As a result, the magnitude of the difference $t_{ji}^{el}(\{A{\bf d}_{{n}\alpha}\})-t_{ji}^{el}(\{{\bf 0}\})$ that appears in Eq.~(\ref{omegaIntensityA})
is smaller for Mode-II than for Mode-I and consequently the IETS intensity $\delta g_{\alpha}$ is seen in Fig.~\ref{Fig2(b)} to be much weaker for Mode-II. For this reason, although the amplitudes of the motion of the sulfur atoms in modes I and II are similar and in both cases the motion is approximately along the molecular axis, {\em Mode-I completely dominates} the calculated IETS spectrum in the phonon energy range that corresponds to the main feature of the experimental IETS phonon histogram\cite{HihathArroyoRubio-BollingerTao08} in Fig.~\ref{Fig2(b)}. 

That the IETS intensity for mode I is much larger than that for mode II can also be understood in terms of the Troisi-Ratner propensity rules\cite{Troisi2006, Troisi2006JCP} for inelastic tunneling intensities. For a very simple model of a perfectly symmetric molecule with symmetric electrodes, Troisi and Ratner \cite{Troisi2006JCP} found the IETS intensity for mode II to be exactly zero as a consequence of symmetry. The IETS intensity for mode II is not (in general) exactly zero in the present model because the electrodes are not symmetric. We note that the electrodes are also not expected to be symmetric in current experimental molecular wire devices such as that of Hihath \emph{et al.}\cite{HihathArroyoRubio-BollingerTao08}

Notice that our theoretical results for phonon Mode-I of PDT molecules bonded to both electrodes in the top site geometry, bonded to both electrodes in the bridge geometry, and bonded to one electrode in the bridge and to the other in the top geometry (the features around 45.5, 40.5, and 42 meV in the theoretical spectra in Fig.~\ref{Fig2(b)}, respectively) are reasonably well converged with respect to increasing gold cluster size. Also the Mode-I phonon energy ranges of the bridge site-bonded and bridge/top site-bonded wires do not overlap and are well separated from the phonon energy range of Mode-I for pure top site bonding. 

Based on these theoretical results,  we propose the following identification of the bonding geometries between the PDT molecules and gold electrodes that gave rise to the main features of the experimental IETS histogram\cite{HihathArroyoRubio-BollingerTao08} that is reproduced in Fig.~\ref{Fig2(b)}:
We identify the main peak in the experimental histogram\cite{HihathArroyoRubio-BollingerTao08} in Fig.~\ref{Fig2(b)} that is centred at $\sim$46~meV as being due to trans-PDT molecules that bonded to both gold electrodes in the top site geometry. Thus we attribute the majority of the counts in the histogram to molecules in the pure top site bonding geometry.
The weaker peak centred near $42$~meV in the experimental histogram matches our predictions for molecules that bond to gold electrodes in the mixed bridge/top site geometry. 
The shoulder of the experimental histogram\cite{HihathArroyoRubio-BollingerTao08}  at lower phonon energies 
centered near $40.5$~meV corresponds to our results for molecules bonding to both electrodes in the bridge site geometry. 
Thus, our results show that IETS experiments can identify experimental realizations of PDT molecular wires in bridge, bridge/top, and top site bonding geometries by measuring the energies of the phonons emitted during electron transport through these systems.

Although the feature near $42$~meV in the experimental histogram is not very prominent, Hihath {\em et al.}\cite{HihathArroyoRubio-BollingerTao08} observed a transition in which the IETS spectrum switched from exhibiting a peak near $42$~meV to one near $46$~meV as the molecular junction was stretched. They therefore argued that the $42$~meV and $46$~meV vibrational modes correspond to distinct conformations of the molecular junction. However, they were not able to identify the conformations that give rise to these modes. Our theoretical results indicate that the switching from the  $42$~meV to the $46$~meV mode corresponds to the junction switching from trans PDT in a mixed top/bridge bonding geometry to trans PDT in a pure top bonding geometry as the junction is stretched. This is consistent with our finding that the pure top bonding geometry corresponds to a larger separation between the gold electrodes than does the top/bridge bonding geometry.

In addition to modes I and II, the theoretical IETS spectrum in Fig.~\ref{Fig2(b)} shows other vibrational modes (labelled ``Mixed Modes")
at lower energies, between $28$ and $36$~meV. The calculated IETS intensities of these modes are weaker than those of Mode-I but in most cases not as weak as those of Mode-II. In the same energy range, the experimental histogram in Fig.~\ref{Fig2(b)}\cite{HihathArroyoRubio-BollingerTao08} shows very few counts. This suggests that the measurements of Hihath {\em et al.}\cite{HihathArroyoRubio-BollingerTao08} were not able to detect features in the inelastic tunneling spectrum having weak intensities and that the intensities of the ``Mixed Modes" between $28$ and $36$~meV in Fig.~\ref{Fig2(b)} (as well as those of Mode-II) were at or below the detection threshold.

\subsubsection{Hollow site bonding}
\label{Hollowresults}     
%
\unitlength=0.01\linewidth
\begin{figure}[] 
\centering
\includegraphics[width=1.0\linewidth]{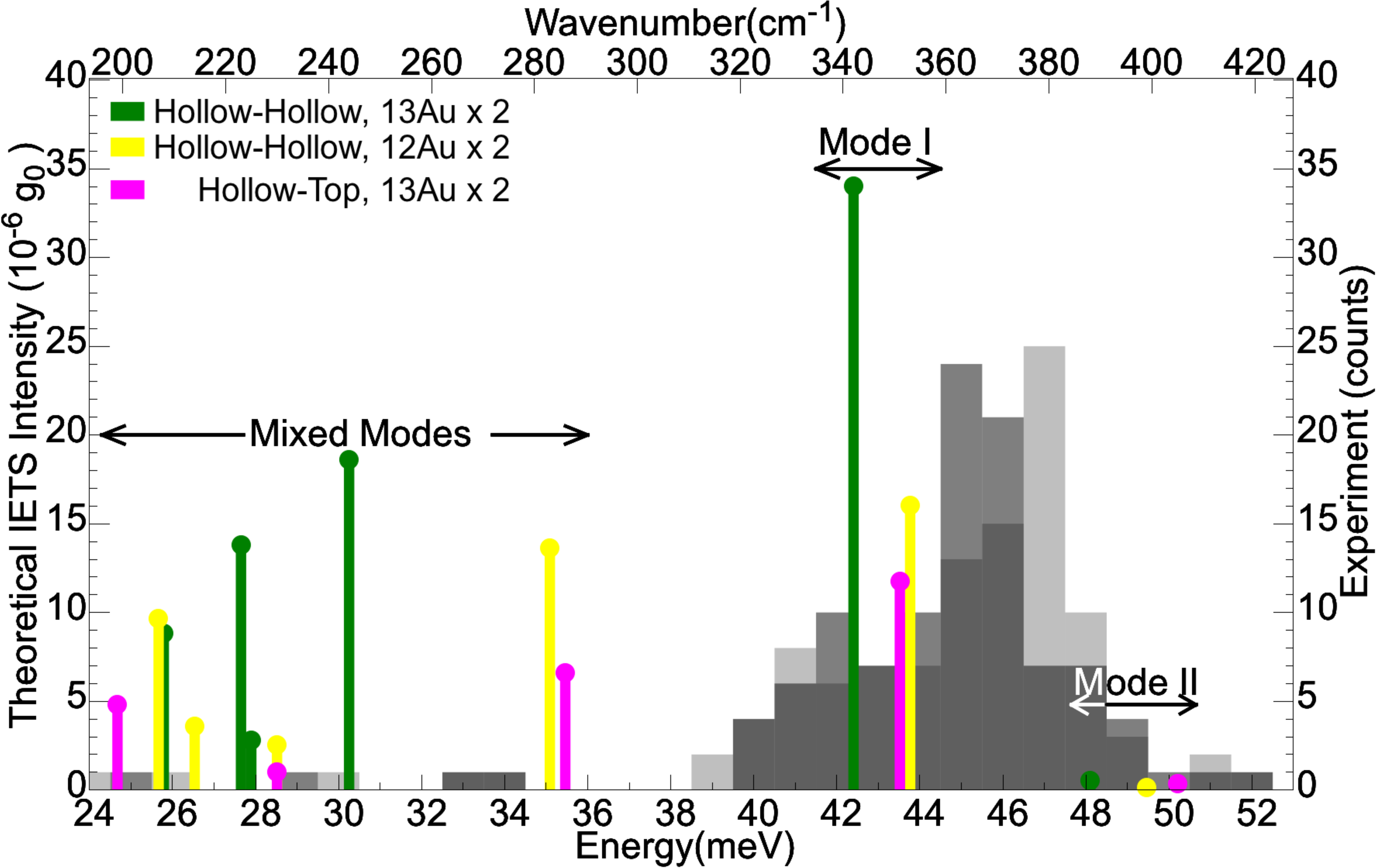}  
\caption{{\color{red}{(Color online.) }}
Calculated IETS intensities (colored) vs. calculated phonon energies for {\em trans}-PDT molecules linking pairs of gold clusters with 12 (yellow) and 13 (green) Au atoms in each cluster for pure hollow site bonding and with 13 Au atoms in each cluster for mixed hollow/top site bonding (magenta).
The experimental IETS phonon mode histogram of Hihath \emph{et al.}\cite {HihathArroyoRubio-BollingerTao08} is shown in (darker, lighter) grey for (positive, negative) bias voltages. 
}
\label{Fig3} 
\end{figure}
For alkanedithiolates with longer chains of (6, 8, or 10) carbon atoms bridging gold electrodes, the low bias conductances have been observed experimentally to drop by a factor of $\sim$4 from the second lowest conductance plateau to the lowest as the molecular junction is stretched.\cite{Li2006} 
It was conjectured \cite{Li2006} that this change in conductance may be due to the molecules switching from hollow site bonding at one electrode and top site bonding at the other (hollow/top), to top site bonding at both gold electrodes.
This conjecture is consistent with the expectation that the distance between gold electrodes should be larger and the conductance smaller for alkanedithiolate molecules bonded to gold electrodes in purely top site geometries than in hollow/top bonding geometries. 
However, whether it is correct or not has remained unclear in the absence of more direct experimental evidence. 

Can such a transition from hollow/top bonding to top site bonding plausibly account for the switch from the $\sim$42 meV phonon mode to the $\sim$46 meV phonon mode that was observed experimentally\cite{HihathArroyoRubio-BollingerTao08} in the IETS spectra of the gold-PDT-gold system?  
As we have already mentioned, in our DFT calculations PDT extended molecules started in hollow site bonded trial geometries in most cases relaxed immediately to bridge-site bonded geometries. This indicates that the hollow site bonding is much more fragile (and thus less likely to be realized experimentally) than bridge or top site bonding. 

Some examples of calculated IETS spectra for PDT molecules hollow site-bonded to one or both gold clusters  are shown in Fig.~\ref{Fig3}.
The calculated phonon energies and IETS intensities for these structures are more sensitive to the sizes of the gold clusters than they are for pure top, pure bridge or mixed bridge/top site bonding. As for the calculated elastic conductances of hollow site bonded structures (see Table \ref{CalculatedConductance}), this greater sensitivity is due to the more open structures of the gold clusters required for hollow site bonding and the resulting greater variability in the {\em details} of the hollow site bonding geometries.  

In Fig.~\ref{Fig3} the energy range in which we find Mode-I phonons for pure hollow site bonding overlaps with the $\sim$42 meV feature of the experimental IETS histogram of Hihath {\em et al.}\cite{HihathArroyoRubio-BollingerTao08}. However, we find the structures with pure hollow site bonding to also support phonon modes with {\em strong} calculated IETS intensities (similar in strength to the Mode-I intensities for the top and bridge site bonded structures in Fig.~\ref{Fig2(b)}) in the mixed mode region below $\sim$38 meV in Fig.~\ref{Fig3} where the experimental IETS histogram shows few or no counts. This suggests that structures with pure hollow site bonding were realized rarely (if at all) in the experiment of Hihath {\em et al.}\cite{HihathArroyoRubio-BollingerTao08}.

For mixed hollow/top bonding the calculated Mode-I phonon energy in Fig.~\ref{Fig3} does not match either the $\sim$42 or the $\sim$46 meV feature in the experimental histogram \cite{HihathArroyoRubio-BollingerTao08} but falls roughly half way between the two. 
Thus, in addition to the formation of bridge/top structures being much more likely (based on our {\em ab initio} simulations) than the formation of hollow/top structures, the bridge/top bonding geometries account for the phonon energy of the 42 meV feature of the experimental\cite{HihathArroyoRubio-BollingerTao08} IETS spectra while the hollow/top bonding geometries do not. Thus it appears unlikely that hollow/top structures played a significant role in the experiment of Hihath {\em et al.}\cite{HihathArroyoRubio-BollingerTao08}.

\subsubsection{Gauche molecular conformations}
\label{gauche}

%
\unitlength=0.01\linewidth
\begin{figure}[ht!] 
\centering

\subfigure[] {   \label{Fig4(a)} 
\includegraphics[width=1.0\linewidth]{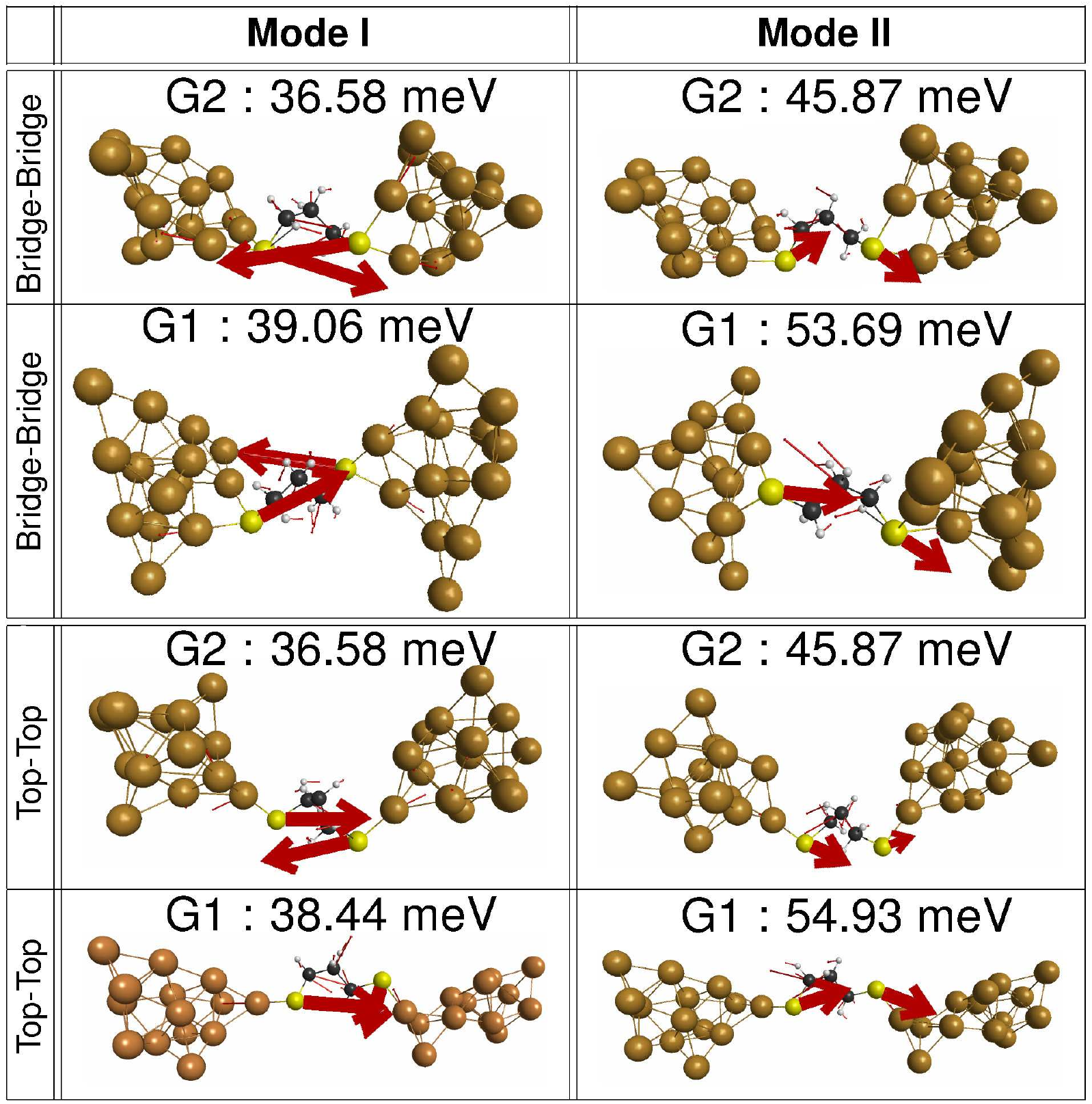}     }
\subfigure[] {   \label{Fig4(b)}
\includegraphics[width=1.0\linewidth]{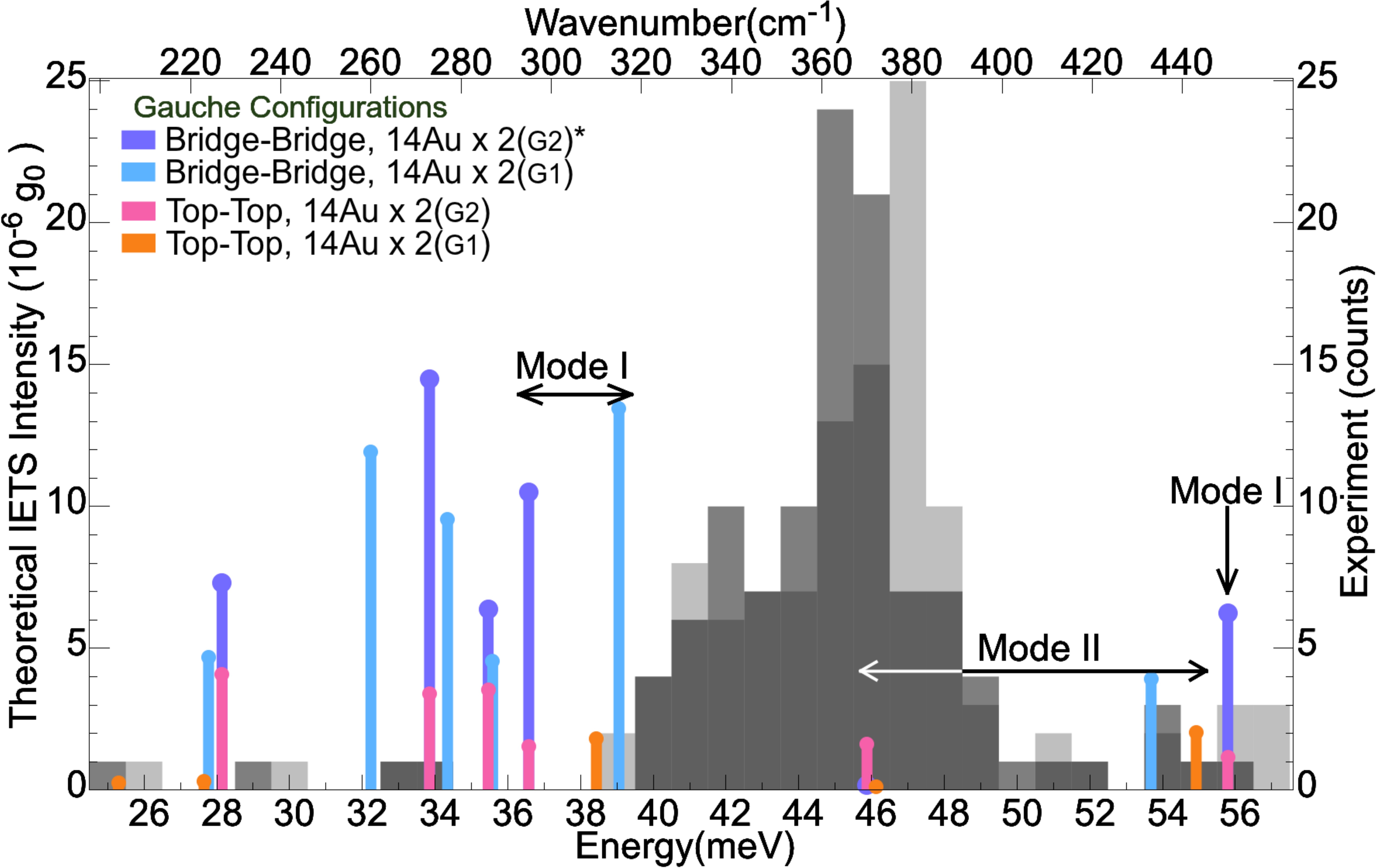}   } 

\caption{{\color{red}{(Color online.) }}
\subref{Fig4(a)} Representative examples of calculated sulfur-related vibrational Mode-I and II of 
 gauche-PDT molecules connecting gold nanoclusters each with 
{\color {red} 14} gold atoms, and the vibrational mode phonon energies. G1 and G2 label molecules with one and two gauche bonds respectively. 
The molecules connect to both gold clusters in pure bridge and pure top site bonding configurations. 
Carbon, hydrogen, sulfur, and gold atoms are black, grey, yellow, and amber, respectively. 
Red arrows show un-normalized atomic displacements. 
\subref{Fig4(b)} IETS intensities of gauche-PDT molecules linking pairs of gold clusters ({\color {red} 14} Au atoms in each cluster) vs. the vibrational mode phonon energy. 
Sulfur atoms bond to the gold in top-top, bridge-bridge site geometries. 
The experimental IETS phonon mode histogram of Hihath \emph{et al.}
\cite {HihathArroyoRubio-BollingerTao08} is shown in (darker, lighter) grey for (positive, negative) bias voltages.
Modes of the types I and II are indicated by arrows. 
* Scaled down by a factor of 10.
}
\label{Fig_gauche} 
\end{figure}

A gauche conformation of an alkanedithiolate molecule can be obtained starting from the trans conformation by rotating part of the molecule through $\sim$$60^{\circ}$ about a C--C or C--S bond. Detection of gauche conformations of PDT molecules adsorbed on gold has been reported\cite{Joo2000}.
Also it has been conjectured \cite{HihathArroyoRubio-BollingerTao08} that the switching that Hihath \emph{et al.} observed\cite{HihathArroyoRubio-BollingerTao08} in their  inelastic tunneling spectra from the 42 meV vibrational mode the to 46 meV mode (as their gold-PDT-gold junction was stretched) might have been due to a change between gauche and trans configurations of the molecular junction. 

We have
investigated theoretically how the presence of gauche bonds would affect the inelastic tunneling spectra of PDT molecules
bridging gold contacts and the possible implications for the experiment.\cite{HihathArroyoRubio-BollingerTao08} 
Representative results of our calculations are presented in Fig \ref{Fig_gauche} (a) and (b) for PDT molecules with one and two gauche bonds connecting pairs of gold clusters in bridge site and top site bonding configurations.  
For the bridge site bonding configurations, the calculated energies of the vibrational modes of the gauche PDT molecules are not  in the vicinity of either 42 or 46 meV. However, the calculated spectra exhibit vibrational modes with very strong IETS intensities at phonon energies between 32 and 38 meV, a range where the experimental histogram\cite{HihathArroyoRubio-BollingerTao08} in Fig.~\ref{Fig_gauche} (b) shows very few counts. Thus it appears unlikely that these structures played a significant role in the experiment of Hihath {\em et al.}\cite{HihathArroyoRubio-BollingerTao08}. 

In Fig.~\ref{Fig_gauche} (b), the PDT molecule with one gauche bond and top site bonding does exhibit a vibrational mode with a phonon energy close to 46~meV. However, this is a type II mode with a low calculated IETS intensity. It is therefore unlikely to have been detected in the IETS experiment.\cite{HihathArroyoRubio-BollingerTao08}
Other calculated vibrational modes of the same top-site bonded gauche structure in the phonon energy range from 27 to 36 meV in Fig.~\ref{Fig_gauche} (b)have somewhat stronger but still weak IETS intensities and occur at energies at which the experimental histogram shows few if any counts. 
The PDT molecule with one gauche bond and top site bonding in Fig.~\ref{Fig_gauche} (b) also exhibits a vibrational mode with a phonon energy close to 46~meV but the calculated 
IETS intensity for this mode is much weaker still.
These results indicate that gauche PDT molecules in top site bonding configurations also appear unlikely to have contributed significantly to the experimental IETS histogram of Hihath {\em et al.}\cite{HihathArroyoRubio-BollingerTao08}. We have also carried out calculations for gauche PDT molecules bonded to the gold electrodes in mixed top-bridge geometries. The results of these calculations (not shown in Fig.~\ref{Fig_gauche}) are qualitatively similar
to those for the pure bridge bonding geometries of gauche molecules discussed above. Namely, these structures were also found not to have vibrational modes with calculated energies in the vicinity of either 42 or 46 meV but they were found to exhibit vibrational modes with very strong IETS intensities at phonon energies between 32 and 38 meV, where the experimental histogram\cite{HihathArroyoRubio-BollingerTao08} shows very few counts. Thus it appears unlikely that these mixed bridge-top bonded gauche molecules played a significant role in the experiment of Hihath {\em et al.}\cite{HihathArroyoRubio-BollingerTao08} 

The low bias {\em elastic} conductance measurements reported by Hihath {\em et al.}\cite{HihathArroyoRubio-BollingerTao08} also provide additional evidence indicating that switching between gauche and trans conformers is unlikely to have been responsible for the experimentally observed transition\cite{HihathArroyoRubio-BollingerTao08} from the structure with the 42 meV vibrational mode to that with the to 46 meV mode, as will be discussed in the following two paragraphs.

We find the calculated sulfur-to-sulfur distances in PDT molecules with gauche bonds to be shorter than those for purely trans PDT molecules. Thus, if the transition from the 42 meV vibrational mode to the 46 meV mode (that occurred as the junction was {\em stretched}) were due to switching between gauche and trans conformations, then the 42 meV structure (being the shorter one) would correspond to the gauche geometry and the 46 meV structure would have the trans geometry. However, the measured conductances of the PDT molecules exhibiting the 42 meV vibrational mode were {\em larger} than those of the PDT molecules exhibiting the 46 meV mode, as can be seen in Fig.4 of Ref. \onlinecite{HihathArroyoRubio-BollingerTao08}. By contrast, recent theoretical work\cite{LiPWTBAE2008,Paulsson.C.F.B.2009} on other alkane-dithiolate molecules bridging gold electrodes
found conductances of pure trans-molecules to be {\em larger} than those of molecules
with gauche bonds, a finding consistent with the results of our calculations of the
low bias conductances for gauche and trans configurations of gold-PDT-gold molecular
wires bonded to the electrodes in the same way; see Table \ref{CalculatedConductance}. These theoretical results suggest that if the switch from the structures with
the $\sim$42 meV phonon to those with the $\sim$46 meV phonon were due 
to a change from a gauche to a trans molecular configuration, then the switch would 
be accompanied by an {\em increase} of the low bias conductance
of the junction,  contrary to what was 
observed experimentally.\cite{HihathArroyoRubio-BollingerTao08} 

By contrast, our calculations show the low bias conductances of gold-{\em trans}-PDT-gold molecular wires in the bridge/top bonding configurations (with the $\sim$42 meV phonon mode) to be {\em higher} than those for {\em trans} molecules in top bonding configurations (with the $\sim$46 meV phonon mode); see Table \ref{CalculatedConductance}. Thus our interpretation of the observed transition from the $\sim$42 meV phonon mode to the $\sim$46 meV phonon mode being due to trans molecules switching from  bridge/top bonding to pure top bonding is  consistent with the experimental low bias conductance data\cite{HihathArroyoRubio-BollingerTao08}.

\subsubsection{Parallel Transport Through Pairs of Molecules }
%
\unitlength=0.01\linewidth
\begin{figure}[h!] 
\centering
\subfigure[] {   \label{Fig5(a)} 
\includegraphics[width=1\linewidth]{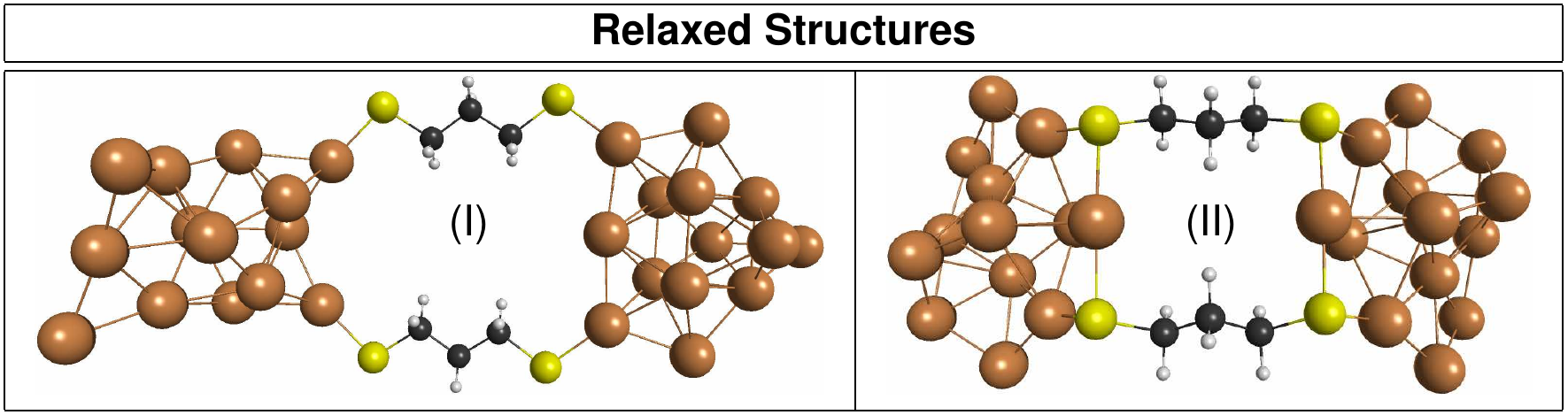}     }
\subfigure[] {   \label{Fig5(b)}
\includegraphics[width=1.01\linewidth]{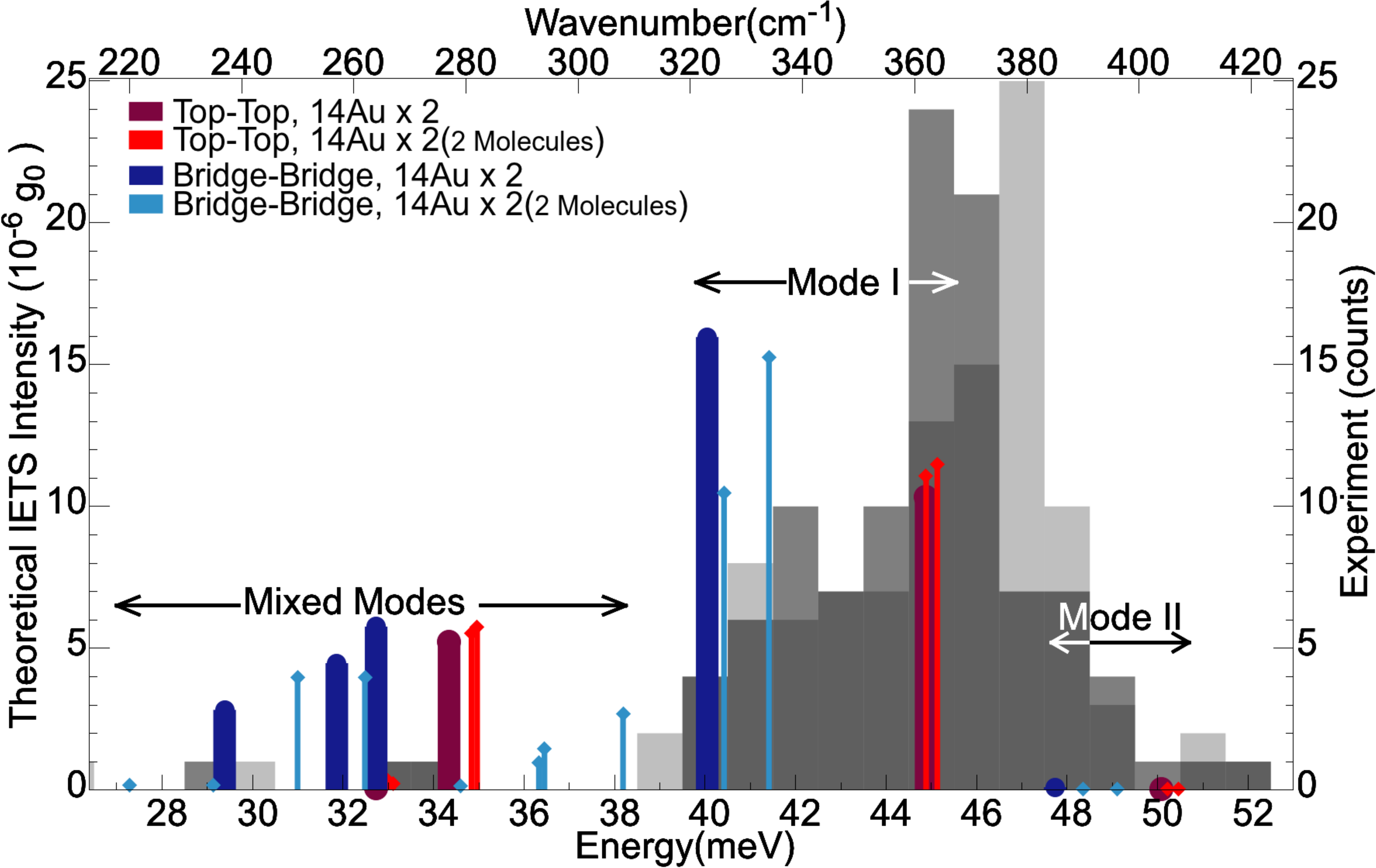}   } 
\caption{{\color{red}{(Color online.) }}
The relaxed geometries and calculated conductances 
for pairs of trans PDT molecules connecting gold clusters in parallel. 
14 gold atoms per cluster. 
(a)-I Top site bonding. 
(a)-II Bridge site bonding. 
Calculated mode frequencies and IETS intensities are shown in (b) for the systems depicted in (a). 
The calculated mode frequencies and IETS intensities for {\em single} molecules similar to those in (a) connecting two gold clusters with the same kinds of molecule-gold bonding geometries are also shown for comparison. 
The experimental phonon mode histogram of Hihath {\em et al.} \cite {HihathArroyoRubio-BollingerTao08} for {\em single} molecules bridging the electrodes is also shown. The darker (lighter) grey represents the positive (negative) bias voltage. 
}
\end{figure}
%
In the preceding discussion we have focussed on cases where a single molecule connects the two electrodes. However in statistical STM break junction experiments, such as that of Hihath \emph{et al.}\cite{HihathArroyoRubio-BollingerTao08}, the electrodes can also be bridged simultaneously by two (or more) molecules  in parallel.  In this Section, we report theoretical results for the inelastic tunneling spectra of some examples of systems of this kind. 

In Figs.~\ref{Fig5(a)} I and II, we show the relaxed geometries for pairs of top site and bridge site bonded trans PDT molecules, respectively, connecting two gold clusters. As can be seen in Table \ref{CalculatedConductance}, 
in each case the calculated low bias elastic conductances of these pairs of molecules are roughly twice that of a single molecule connecting the contacts and bonded to them in a similar configuration. 
This means that the molecules in these systems are behaving as nearly independent parallel quantum conductors in a way reminiscent of parallel semiconductor quantum wires\cite{Castano90} connecting a pair of electron reservoirs. Similar, roughly additive, behavior of the low bias conductances of other molecules bridging pairs of metal electrodes in parallel has also been found in previous theoretical studies, for not too small intermolecular separations \cite{Yaliraki98, Magoga99, Lang2000, Lagerqvist2004, Liu2005, Geng2005, Long2007,
Wang2010, Reuter2011}. 

The calculated inelastic tunneling spectra for these pairs of molecular wires are shown in Fig.~\ref{Fig5(b)}. The inelastic tunneling spectra for single molecules connecting the two electrodes with similar bonding geometries are also shown for comparison. The Mode I inelastic tunneling spectra for the pairs of molecules are very similar to the corresponding spectra of the individual molecules apart from small phonon energy shifts and splittings, indicative of weak effective vibrational coupling between the molecules. However, for the pair of bridge site bonded molecules, the two Mode I features have noticeably differing calculated IETS intensities. 

We also studied pairs of molecules with even smaller spatial separations than those of the molecules in Fig. \ref{Fig5(a)}. Namely, we considered pairs of top site bonded trans PDT molecules binding to nearest neighbor atoms of both gold clusters, as distinct from the second neighbor gold atoms to which the molecules bind in Fig. \ref{Fig5(a)}-I. As expected based on previous studies \cite{Yaliraki98, Magoga99, Lang2000, Lagerqvist2004, Liu2005, Geng2005, Long2007,
Wang2010, Reuter2011}, the calculated low bias conductances of these systems differed more from twice the conductance of a single molecule connecting the gold clusters than was the case for the more widely spaced molecules in Fig. \ref{Fig5(a)}. The splittings of the Mode-I phonon energies for these more closely spaced molecules were also considerably larger. However, we found that relaxed structures with such closely spaced molecules could only be generated for the smaller gold clusters that we studied; for the larger clusters DFT relaxations resulted instead in structures similar to those in Fig. \ref{Fig5(a)}. This suggests that structures with such closely spaced pairs of molecules are unlikely to be realized in experiments such as those of Hihath \emph{et al.}\cite{HihathArroyoRubio-BollingerTao08}.

Experimental inelastic tunneling spectra for molecules bridging the contacts in parallel were not reported in Ref.~\onlinecite{HihathArroyoRubio-BollingerTao08}. Thus comparison between these theoretical results and experiment, while of considerable interest, is not possible at present.

\section{Conclusions} \label{Conclusions}

We have calculated the relaxed geometries, zero bias conductances, and inelastic tunneling spectra of trans and gauche PDT molecules bonded to pairs of gold clusters in various ways and compared our results with the experimental STM break junction transport data of Hihath \emph{et al.} \cite{HihathArroyoRubio-BollingerTao08} Although the gold clusters that represent the macroscopic electrodes in our calculations were not large (up to 14 gold atoms per cluster) they were sufficiently large for accurate modeling of the properties of the vibrational modes that we studied (as is discussed in Section \ref {TBHSresults}) and our calculated inelastic tunneling spectra converged well with increasing size of the gold clusters. Thus, we expect our findings to be applicable to molecular wires bridging macroscopic gold electrodes. Consistent with this, our theoretical results agree well with the experimental data of Hihath \emph{et al.} \cite{HihathArroyoRubio-BollingerTao08} 
We conclude that IETS when combined with such calculations is able to reveal detailed, previously inaccessible, information about the atomic scale structures of the molecule-metal interfaces and thus resolve the long standing ``contact problem" of single-molecule nanoelectronics. 

The success of our approach rests on the fact
that {\em ab initio} density functional theory calculations of vibrational modes and
their frequencies are believed to be reasonably accurate. The reason for this is that in the Born-Oppenheimer
approximation they are electronic {\em ground state total energy}
calculations and density functional theory has been designed and carefully optimized for calculations of the total ground state energies of interacting electron systems.\cite{review2010} While the results of {\em transport} calculations
for molecular nanowires are well known to be less accurate, due in part to the well known difficulty 
of estimating the alignment of the molecular energy levels relative to the Fermi energy of the electrodes
\cite{review2010},
we rely on transport calculations {\em only for qualitative information}
such as the identification of the phonon mode in a
particular frequency range that has the largest IETS intensity. Our
identification of this mode is also supported by physical  reasoning. 
 
We have definitively identified particular realizations of gold-propanedithiolate-gold molecular wires in a recent 
experiment\cite{HihathArroyoRubio-BollingerTao08} in which a trans PDT molecule bonded to a single
gold atom of each electrode. We showed that this bonding geometry and molecular conformation was the one realized most frequently in the experiment.\cite{HihathArroyoRubio-BollingerTao08} Our theoretical results also showed that the switching from the  $42$~meV to the $46$~meV mode observed in the experiment as the molecular wire was stretched\cite{HihathArroyoRubio-BollingerTao08} corresponds to the junction switching from trans PDT in a mixed top/bridge bonding geometry to trans PDT in a pure top bonding geometry.

Our results also showed that PDT molecular wires with gauche conformations connecting gold electrodes should have inelastic tunneling spectra that differ significantly from those with trans conformations. Comparing our results with the inelastic transport data of Hihath \emph{et al.} \cite{HihathArroyoRubio-BollingerTao08}, we concluded that gauche conformations were realized at most rarely in their experiment\cite{HihathArroyoRubio-BollingerTao08}.

We also found that if a pair of PDT molecules connects two gold electrodes in a parallel geometry the coupling between the molecules should affect the low bias elastic and inelastic transport characteristics of the system only weakly, unless the molecules are extremely close together, for example, if they bond in a top site geometry to adjacent gold atoms of both electrodes. Structures with such closely spaced molecules were found to be unstable for the larger gold clusters in our DFT calculations.

\section{Acknowledgments}
This research was supported by CIFAR, NSERC, Westgrid, and Sharcnet. 
We thank J. Hihath, N. J. Tao, E. Emberly, N. R. Branda, V. E. Williams, and A. Saffarzadeh 
for helpful discussions, and J. Hihath and N. J. Tao for providing to us their experimental data 
in digital form.


\end{document}